\documentclass[aps,prb,twocolumn,notitlepage,showpacs,superscriptaddress]{revtex4-2}
\usepackage{graphicx}
\usepackage{amsmath,amssymb}
\usepackage[T1]{fontenc}
\usepackage{lmodern}
\usepackage[utf8]{inputenc}
\usepackage{natbib}
\usepackage{hyperref}
\usepackage{color}
\hypersetup{colorlinks=true,citecolor={blue},linkcolor={blue},urlcolor={blue}}
\usepackage{units}
\usepackage[english]{babel}
\usepackage[normalem]{ulem}
\usepackage{tcolorbox}

\begin{document}

\title{Screening nearest-neighbor interactions in networks of exciton-polariton condensates through spin-orbit coupling}

\author{Denis Aristov}
\affiliation{Hybrid Photonics Laboratory, Skolkovo Institute of Science and Technology, Territory of Innovation Center Skolkovo, 6 Bolshoy Boulevard 30, building 1, 121205 Moscow, Russia}

\author{Helgi Sigurdsson}
\email[H. Sigurdsson ]{helg@hi.is}
\affiliation{Science Institute, University of Iceland, Dunhagi 3, IS-107, Reykjavik, Iceland}
\affiliation{Department of Physics and Astronomy, University of Southampton, Southampton SO17 1BJ, United Kingdom}

\author{Pavlos G. Lagoudakis}
\affiliation{Hybrid Photonics Laboratory, Skolkovo Institute of Science and Technology, Territory of Innovation Center Skolkovo, 6 Bolshoy Boulevard 30, building 1, 121205 Moscow, Russia}
\affiliation{Department of Physics and Astronomy, University of Southampton, Southampton SO17 1BJ, United Kingdom}

\date{\today}

\begin{abstract}
We study the modification of the spatial coupling parameter between interacting ballistic exciton-polariton condensates in the presence of photonic spin orbit coupling appearing from TE-TM splitting in planar semiconductor microcavities. We propose a strategy to make the coupling strength between next-nearest-neighbours stronger than between nearest-neighbour, which inverts the conventional idea of the spatial coupling hierarchy between sites. Our strategy relies on the dominantly populated high-momentum components in the ballistic condensates which, in the presence of TE-TM splitting, lead to rapid radial precession of the polariton pseudospin. As a consequence, condensate pairs experience distance-periodic screening of their interaction strength, severely modifying their synchronization and condensation threshold solutions.
\end{abstract}

\maketitle

\section{Introduction}
Building large-scale optically programmable systems of interacting quantum gases permits exploration of spatially extensive many-body physics such as buildup of correlations, many-body localization, or critical phase transitions across systems covering lattices~\cite{Morsch_RMP2006}, glassy networks~\cite{Fallani_PRL2007}, quasicrystalline structures~\cite{Freedman_Nat2006, Viebahn_PRL2019} and so on. From a more practical viewpoint, programmable condensed matter systems are favorable for developing simulators and natural computing strategies in the semiclassical regime~\cite{Berloff_NatMat2017, Gershenzon_Nanopho2020, Ballarini_NanoLett2020, Kalinin_NanoPho2020, Kiraly_NatNanoTech2021} or the deep quantum regime~\cite{Georgescu_RMP2014}. For these purposes, intense research effort has been focused on advancing optical lattices for cold atom ensembles ~\cite{Morsch_RMP2006}, and---quite recently---single Rydberg atoms per lattice site using optical tweezers~\cite{Browaeys_NatRevPhys2020}, thermo-optically imprinted photon condensate potentials~\cite{Dung_NatPho2017}, and exploiting the strong repulsive exciton-exciton Coulomb interaction in nonresonantly driven exciton-polariton condensates~\cite{Ohadi_PRB2018, Zhang_NanoScale2018, Pickup_NatComm2020, Pieczarka_Optica2021}. Naturally, from the hierarchical overlap between different localized lattice-site wavefunctions, the nearest-neighbour (NN) interactions tend to be always stronger than those of next-nearest-neighbours (NNN) and strategies to overcome such a basic spatial constraint in condensed matter systems are scarce, and thus it remains mostly unexplored.

Here, we propose a method in which NNN interactions are made stronger than NN interactions in a system of ballistic exciton-polariton ({\it polariton} hereafter) condensates with photonic spin-orbit-coupling (SOC). Polaritons are a mixture of both light and matter arising in the strong coupling regime between quantum well excitons and cavity photons in semiconductor microcavities~\cite{kavokin_microcavities_2007}. Being composite bosons, they can undergo a power-driven nonequilibrium phase transition into a highly coherent many-body state referred as a polariton condensate~\cite{Carusotto_RMP2013}. Polaritons possess a two-component pseudospin structure (or just {\it spin}) along the growth axis of the microcavity that is explicitly connected to the photon circular polarization and thus measurable through the polariton photoluminescence. Being a driven-dissipative nonequilibrium system, polariton condensates can exist at energies far above the ground state~\cite{Tosi_NatPhys2012} and form expanding fluids of light with large momentum. Such ballistic condensates can be generated using tightly focused~\cite{Wertz_PRL2012} or specially shaped optical excitation beams~\cite{Assmann_PRB2012}, leading to robust synchronization between condensate pairs, recorded over 100 $\mu$m separation distance~\cite{Topfer_ComPhys2020} (roughly 50 times larger than the condensate width).

In planar cavities the polarization of the photons is degenerate at normal incidence $\mathbf{k}_\parallel = 0$ except in the presence of disorder or strain. However, at oblique angles $\mathbf{k}_\parallel \neq 0$ photons experience different transmission and reflection from the cavity which depends on their polarization. This leads to splitting between the transverse electric (TE) and transverse magnetic (TM) modes, referred as simply TE-TM splitting~\cite{Panzarini_PRB1999}. This forms an effective SOC magnetic ﬁeld which lies in the plane of the cavity with a double winding in reciprocal space [see Fig.~\ref{fig1}(a)] and the eigenmodes of the system are linearly polarized parallel or perpendicular to their propagation direction. Since photons impart their polarization structure directly onto the polaritons, various phenomena have been proposed and observed such as the optical spin Hall effect~\cite{Kavokin_PRL2005, leyder_observation_2007}, polariton topological physics~\cite{Nalitov_PRL2015, Karzig_PRX2015, Solnyshkov_OptMatExp2021}, reduction to other SOC types~\cite{Whittaker_NatPho2021}, baby skyrmions~\cite{Cilibrizzi_PRB2016}, and much more.

Combining the recent advancements in creating ballistic polariton condensates of high radial momentum~\cite{Topfer_ComPhys2020, Topfer_Optica2021}, and the familiar photonic TE-TM splitting, we demonstrate distance-periodic screening of spatial interactions in polariton condensate networks. Our method presents a new all-optical tool to reversibly engineer spatial interactions in extended nonequilibrium coherent quantum gases, and offers perspectives towards exploring many-body physics with unusual coupling hierarchies in the optical regime. As an example, the method can be used to design non-planar graph topologies of condensates which opens pathways toward simulation of complex XY phases~\cite{Berloff_NatMat2017} (in analogy to the NP-complete non-planar Ising model~\cite{Barahona_IOP1982}) and unexplored regimes of synchronicity in large-scale nonlinear oscillator networks. Moreover, our method can also work with the more familiar Rashba or Dresselhaus SOC in condensed matter systems.

\section{Spinor polariton model}
The spinor polariton condensate macroscopic wavefunction $\Psi(\mathbf{r},t) = (\psi_+, \psi_-)^\text{T}$ is modeled using the well known and accepted generalized Gross-Pitaevskii equation coupled to a rate equation describing the laser-driven exciton reservoir $\mathbf{X}(\mathbf{r},t) = (X_+,X_-)^\text{T}$~\cite{Carusotto_RMP2013},
\begin{align} \notag
    & i\frac{\partial \psi_{\pm}}{\partial t}  = \bigg[-\frac{\hbar\nabla^{2}}{2m} + \frac{i}{2}\left(R X_\pm - \gamma\right) + \alpha |\psi_{\pm}|^{2} \\ 
    & + G \left(X_\pm + \frac{P_\pm(\mathbf{r})}{W} \right) \bigg]\psi_{\pm} + \Delta_{LT} \left( \frac{\partial}{\partial_x} \mp i \frac{\partial}{\partial_y}\right)^2 \psi_\mp,
    \label{eq.GPE} \\
    & \frac{\partial X_{\pm}}{\partial t}  = -\left(\Gamma  + R|\psi_{\pm}|^2\right)X_{\pm} + P_\pm(\mathbf{r}).
    \label{eq.Res}
\end{align}
Here, $m$ is the polariton mass, $\gamma^{-1}$ the polariton lifetime, $G = 2 g |\chi|^2$ and $\alpha = g |\chi|^4$ are the same spin polariton-reservoir and polariton-polariton interaction strengths, respectively, $g$ is the exciton-exciton Coulomb interaction strength, $|\chi|^2$ is the excitonic Hopfield fraction of the polariton, $R$ is the scattering rate of reservoir excitons into the condensate, $\Gamma$ is the reservoir decay rate, $W$ quantifies additional blueshift coming from a background of high-energy and dark excitons generated by the nonresonant continuous-wave pump $\mathbf{P}(\mathbf{r}) = (P_+,P_-)^\text{T}$, whose $\pm$ components correspond to incident $\sigma^\pm$ circular polarized light, respectively. The parameters are based typical GaAs microcavity properties and fitting to previous experiments~\cite{Topfer_ComPhys2020}: $m = 5 \times 10^{-5} m_0$ where $m_0$ is the free electron mass; $\gamma^{-1} = 5.5$ ps; $|\chi|^{2}=0.4$ since our cavity is negatively detuned; $\hbar g = 0.5 \, \mu\mathrm{eV\,\mu m^{2}}$; $R=3.2g$; and $W = \Gamma = \gamma$. 

\begin{figure}
    \centering
    \includegraphics[width = \linewidth]{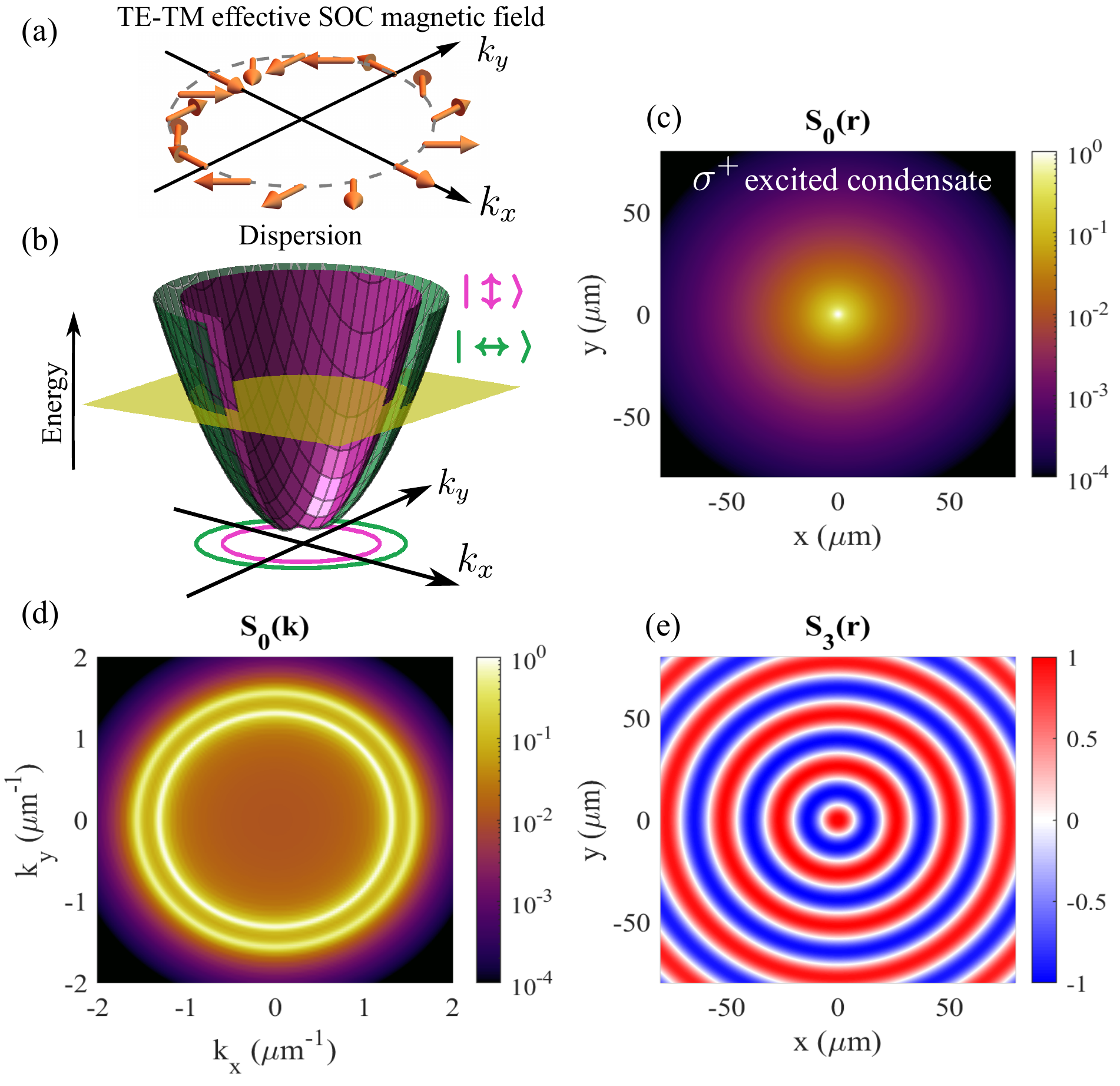}
    \caption{(a) Schematic of the effective SOC magnetic field from the TE-TM splitting on a momentum-space circle. (b) Schematic of the polariton dispersion forming two concentric parabolas. Circles denote the intersection of the isoenergy plane. (c) Condensate steady state real space density for a single pump above threshold centered at the origin showing an expanding cloud of polaritons. (d) Condensate steady state reciprocal-space density showing the two illuminated isoenergy dispersion rings due to the TE-TM splitting. Here we have set $\Delta_{LT} = 0.2$ ps$^{-1}$ $\mu$m$^2$. (e) Corresponding real space $S_3$ component showing the radial precession of the pseudospin.}
    \label{fig1}
\end{figure}

The last term in Eq.~\eqref{eq.GPE} describes the cavity photon TE-TM splitting~\cite{ Kavokin_PRL2005} which lifts the spin-degeneracy of the polariton dispersion. It can be visualized as an effective spin-orbit coupling in-plane magnetic field which does a double angle rotation in momentum space as shown in Fig.~\ref{fig1}(a). The eigenmodes belonging to the co-centric double parabola [see Fig.~\ref{fig1}(b)] are linearly polarized parallel or perpendicular to their propagation direction, whose different effective masses $m_{\text{TE},\text{TM}}$ are given by,
\begin{equation}
    \Delta_{LT} = \frac{\hbar}{4}\left( \frac{1}{m_\text{TM}} - \frac{1}{m_\text{TE}} \right).
\end{equation}

To keep things simple, spin relaxation between the reservoir excitons~\cite{Vina_JPCM1999, Pickup_PRB2021} is not included in the main analysis since it only introduces trivial depolarization to the reservoir steady state. This immediately implies that one must work sufficiently close to threshold since driving the system at higher pumping powers might lead to condensation of the opposite spin component at the pump spots. A brief discussion on this aspect is given in Appendix~\ref{app1}. We have also neglected cross-spin interactions since recent experiments have found its strength to be only few percent of the same-spin interaction strength~\cite{Bieganska_PRL2021}.

We consider the simplest method of generating ballistic polariton condensates using nonresonant circularly polarized optical excitation profile described with a superposition of Gaussians forming the pump spots,
\begin{equation}
    P_+(\mathbf{r}) = P_{0}  \sum_{n} e^{-|\mathbf{r} - \mathbf{r}_n|^2 / 2w^2}, \qquad P_- = 0.
    \label{eq.pump}
\end{equation}
Throughout the study we will use Gaussians with $2 \ \mu$m full-width-at-half-maximum, representing tightly focused excitation beams.

It is instructive to write out explicitly the form of the complex polariton potential in Eq.~\eqref{eq.GPE} for the case of weak nonlinearities $|\psi_\pm|^2 \simeq 0$, representing the system being below or close to the condensation threshold. The steady state of the reservoir is then $X_\pm \approx P_\pm/\Gamma$ and we can write,
\begin{equation}
  V_\pm(\mathbf{r}) \approx P_\pm(\mathbf{r}) \left[ \frac{i R  }{2 \Gamma} + G \left(\frac{1}{\Gamma} + \frac{1}{W} \right) \right].
  \label{eq.pot}
\end{equation}
The condensation threshold of polaritons can be identified as the point where a single frequency component of the linearized Eq.~\eqref{eq.GPE} crosses from negative to positive imaginary value by ramping the power density $P_0 > P_{0,\text{thr}}$. Subsequently, a transient phase occurs where $|\Psi|$ grows in time until it reaches saturation through the reservoir $\mathbf{X}$ gain clamping.

\begin{figure*}
    \centering
    \includegraphics[width = \linewidth]{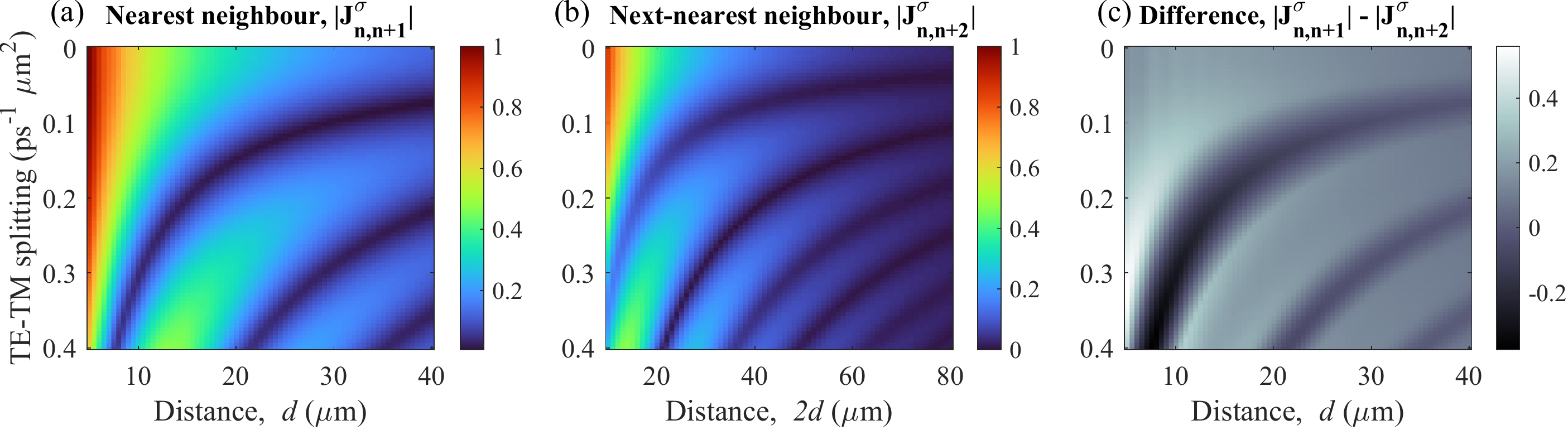}
    \caption{Amplitude from the overlap integral~\eqref{eq.overlap_int} between  nearest neighbour (a) and next-nearest neighbours (b) as a function of TE-TM splitting strength and distance $d$ between neighbours. The panels are normalized with respect to each other. (c) Difference between the two integrals.}
    \label{fig2}
\end{figure*}

Since polaritons are two-component spinors they can be associated with a three-dimensional pseudospin which, in terms of the emitted cavity photoluminescence, is analogous to the Stokes vector of light,
\begin{equation}
    \mathbf{S} = 
    \begin{pmatrix} S_1 \\ S_2 \\ S_3 \end{pmatrix} = \frac{\Psi^\dagger \boldsymbol{\hat{\sigma}} \Psi}{\Psi^\dagger \Psi}.
\end{equation}
Here, $\boldsymbol{\hat{\sigma}}$ is the Pauli operator vector. The pseudospin can be directly measured in both the near-field and far-field through standard polarimetry detection setup in experiments. It is also good to define the fourth pseudospin (Stokes) component which represents the total density in the system,
\begin{equation}
    S_0 = \Psi^\dagger \Psi = |\psi_+|^2 + |\psi_-|^2.
\end{equation}
We will refer to the pseudospin components in real space and reciprocal (momentum) space through their arguments, $S_n(\mathbf{r})$ and $S_n(\mathbf{k})$, respectively.
\section{Results}
In this section we will demonstrate that the spatial interactions between two ballistic condensates can be manipulated using TE-TM splitting. We start by characterizing the spin properties of the single condensate. We then work our way up and analyze the case of two interacting condensates, three condensates in a chain, and a square of four condensates.

\subsection{Single condensate}
Here, a single Gaussian pump spot is used and driven above threshold $P_0 > P_{0,\text{thr}}$. The condensate wave function $\Psi$ and reservoir $\mathbf{X}$ quickly converge to the steady state solution presented in Fig.~\ref{fig1} displaying the monotonically radially decaying envelope of the condensate~\cite{Wouters_PRB2008} centered at the pump spot location at the origin [Fig.~\ref{fig1}(c)]. Since the pump is circularly polarized the spin-up polaritons are dominantly excited at the origin which can be seen from the red central area in the $S_3(\mathbf{r})$ in Fig.~\ref{fig1}(e). As the condensate expands from the center the TE-TM splitting causes a precession of the pseudospin~\cite{Kamman_PRL2012, Cilibrizzi_PRB2016} which alternates from spin-up (red) to spin-down (blue) radially with a period $\xi = 2 \pi / \Delta_k$ where $\Delta_k = |\sqrt{2 m_\text{TE} \omega_c / \hbar} - \sqrt{2 m_\text{TM} \omega_c / \hbar}|$ is the spacing between the excited reciprocal space circles [Fig.~\ref{fig1}(d)] and $\hbar \omega_c$ is the energy of the condensate denoted with the yellow plane in Fig.~\ref{fig1}(b). 

\subsection{Two condensates}
The precession of the polariton condensate $S_3(\mathbf{r})$ component as waves propagate away from the pump spot [see Fig.~\ref{fig1}(e)] leads to periodic variations in the overlap integral between neighbouring condensates depending on their separation distance $\mathbf{d} = \mathbf{r}_n - \mathbf{r}_{n+1}$. The overlap integral quantifies the amount of coupling between any two condensates which can be written,
\begin{equation} \label{eq.overlap_int}
    J_{n,m}^{\sigma} = \int \psi^*_{ \sigma}(\mathbf{r} - \mathbf{r}_n) V_\sigma(\mathbf{r}) \psi_{\sigma}(\mathbf{r} - \mathbf{r}_m) \, d\mathbf{r},
\end{equation}
where we have adopted a shorthand notation $\sigma \in \{ \pm\}$. Because $V_\sigma(\mathbf{r})$ is complex, the coupling $J_{n,m}^\sigma$ is non-Hermitian and represents both renormalized energies and linewidths of the polariton modes. Specifically, if the TE-TM splitting is chosen such that $d = \xi/2$ it becomes clear that spin-up(down) waves propagating from $\mathbf{r_n}$ will be mostly converted into spin-down(up) waves at the NN location $\mathbf{r}_{n+1}$. The polariton spin then completes a full revolution upon reaching the NNN at $\mathbf{r}_{n+2}$. In this kind of scenario, one can conjecture the following inequality,
\begin{equation}
  |J_{n,n+1}^{\sigma}| \lesssim  |J_{n,n+2}^{\sigma}|,
\end{equation}
exemplifying a system where NNN interactions are greater or comparable to the NN interactions.

The above inequality can be tested by straightforwardly calculating Eq.~\eqref{eq.overlap_int} as a function of TE-TM splitting and separation distance using the numerically obtained wavefunction profile in Fig.~\ref{fig1}. We consider a line of three pumps written,
\begin{equation}
    P_+(\mathbf{r}) \propto e^{-|\mathbf{r} - d \mathbf{\hat{x}}|^2 / 2w^2} +
    e^{-r^2 / 2w^2} + 
    e^{-|\mathbf{r} + d \mathbf{\hat{x}}|^2 / 2w^2},
\end{equation}
while $P_-(\mathbf{r}) = 0$. Using Eq.~\eqref{eq.pot} to obtain $V_\sigma(\mathbf{r})$, and using the condensate wavefunction from Fig.~\ref{fig1}(c) to approximate $\psi_\sigma(\mathbf{r} - \mathbf{r}_n)$, we can calculate the overlap integral~\eqref{eq.overlap_int}.
The results are shown in Fig.~\ref{fig2} where indeed clear regimes of coupling strength inversion between NNs and NNNs can be identified [dark regions in Fig.~\ref{fig2}(c)].

At this point we would like to stress that the periodic radial precession of the $S_3$ pseudospin component shown in Fig.~\ref{fig1}(e), and the results in Fig.~\ref{fig2}, are not exclusive to the double winding SOC field coming from TE-TM splitting. The results of this study can also be easily constructed with the high-momentum condensates using the more familiar single winding Rashba $\hat{H}_R = \hat{\sigma}_x k_x - \sigma_y k_y$ or Dresselhaus $\hat{H}_D = \hat{\sigma}_x k_y - \sigma_y k_x$ SOC fields. This fact underlines the applicability of our proposed method to engineer interactions across other condensed matter systems~\cite{Galitski_Nature2013}. However, the advantage of exciton-polariton condensates are their high-momentum (ballistic) nature which gives extreme propagation distances, with subsequent long interaction and coherence lengths~\cite{Topfer_ComPhys2020, Topfer_Optica2021}. 

\begin{figure}
    \centering
    \includegraphics[width = \linewidth]{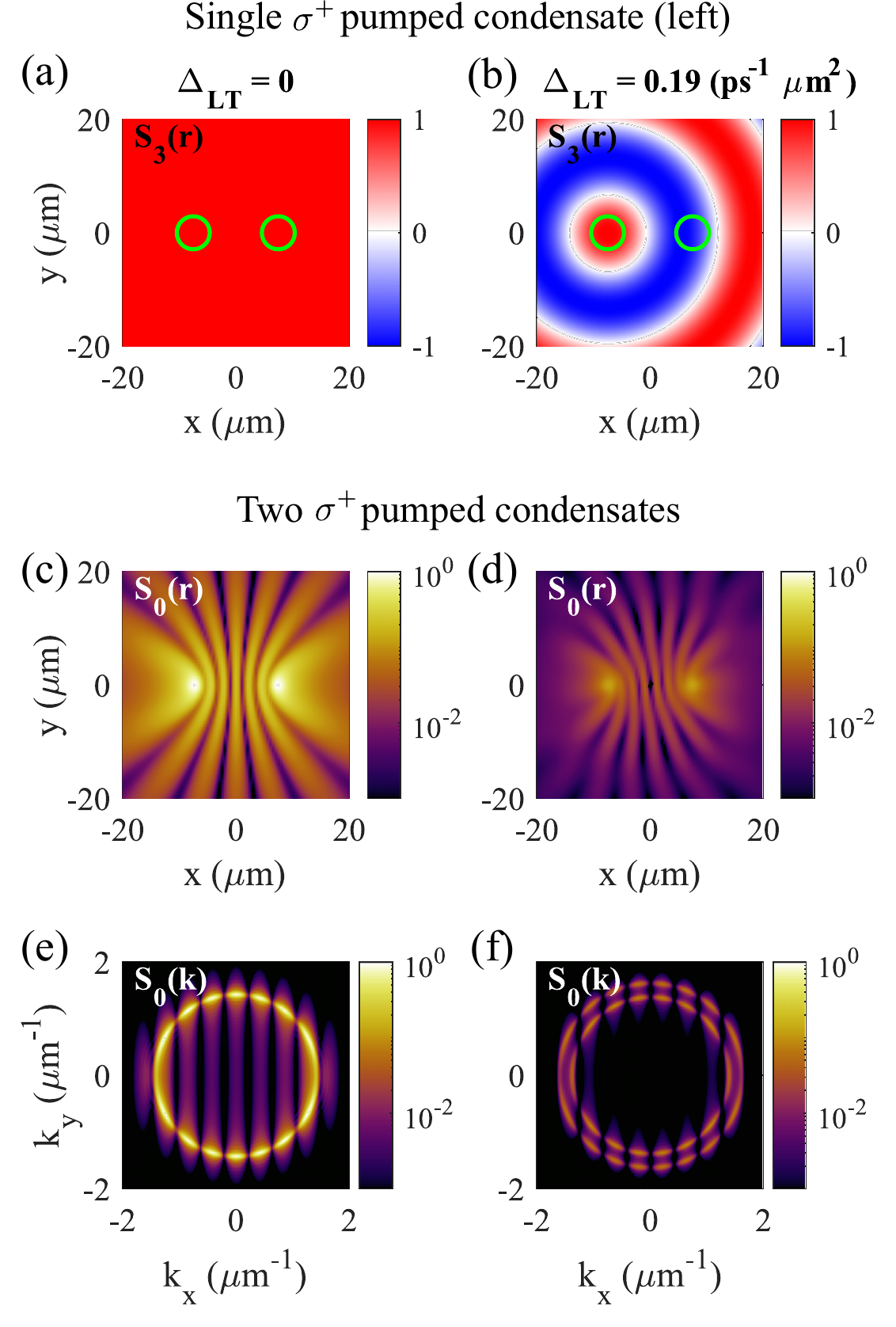}
    \caption{Real space $S_3$ Stokes component for only the left spot pumped above threshold without TE-TM splitting (a) and with TE-TM splitting (b). Circles denote the location of the pumps used. (c,d) Normalized total condensate steady state density in real space and (e,f) momentum space for two co-circularly polarized pumps driven above threshold separated by $d = 15$ $\mu$m, and for the two corresponding different values of TE-TM splitting.}
    \label{fig3}
\end{figure}

To verify the calculation presented in Fig.~\ref{fig2} we simulate a two-condensate system (i.e., a polariton dyad~\cite{Topfer_ComPhys2020}) by pumping with two Gaussians separated by a distance $d$. The value of $d$ is chosen so to favor a single energy state (i.e., stationary state or \emph{fixed point} solution) since it was recently established that the polariton dyad has a large family of both stationary and non-stationary solutions depending on distance~\cite{Topfer_ComPhys2020}. We have chosen a distance of $d=15$ $\mu$m which places each condensate in the half-period of the $S_3(\mathbf{r})$ pattern, minimizing the overlap. This is depicted in Fig.~\ref{fig3}(b) where the green circles denote the locations of the condensate pumps and only the left condensate is being pumped above threshold. One can observe that the position of the right condensate pump is comfortably placed in the half-period blue region where all polaritons flowing from the neighbor have nearly all converted to the opposite spin component. We point out that if the TE-TM splitting is absent then the condensate becomes homogeneously polarized [see Fig.~\ref{fig3}(a)].

Results when the two pump spots are pumped above threshold are shown in Fig.~\ref{fig3}(c)-(f) where we show the real- and momentum space density distributions of the steady state condensate for two different values of TE-TM splitting [Fig.~\ref{fig3}(c,e) and~\ref{fig3}(d,f), respectively]. In each case the condensate dyad converges into a state of definite parity observed from the clear interference fringes in both real- and momentum space.

Figures~\ref{fig3}(d) and~\ref{fig3}(f) are normalized with respect to Figs.~\ref{fig3}(c) and~\ref{fig3}(e), respectively. As expected, the decreased coupling strength due to the presence of TE-TM splitting results in a dramatically weaker condensate density. Another interesting feature is the change in parity between the two cases. Figure~\ref{fig3}(e) shows a bright antinode at $k_x=0$ whereas Fig.~\ref{fig3}(f) shows a dark nodal line. Therefore, the TE-TM splitting not only weakens the coupling but also changes the interference condition between condensates due to the altered dispersion.

Another notable feature in the condensate in the presence of TE-TM splitting is the warped real space intensity $S_0(\mathbf{r})$ in Fig.~\ref{fig3}(d). This chiral pattern stems from the SOC which converts propagating polariton spins into the opposite spin component which also obtains orbital angular momentum to conserve the total angular momentum~\cite{Manni_PRB2011}. As a result, the pumped $|\psi_+|^2$ polaritons form a mirror symmetric standing wave pattern between the pump spots~\cite{Topfer_ComPhys2020} whereas spin-down polaritons $|\psi_-|^2$ obtain a warped density distribution of definite chirality (see Appendix~\ref{app2}). Note that, technically, these are not true standing waves but ``leaky'' standing waves due to transverse losses. This chiral pattern is observed in all results with TE-TM splitting but is not discussed further.

The weakened coupling between the two condensates can be quantified using two different measures. First, the total particle number is expected to drop when the condensate coupling strength is small because of lessened cooperativity between the condensates. This measure is written,
\begin{equation}
N = \int S_0 \, d\mathbf{r}.  
\end{equation}
Second, the degree of coherence between the condensates in the presence of random noise will be weakened when their coupling strength is small. The mutual coherence measure can be calculated using the standard definition for first order correlation function (or \emph{coherence}) for vectorial fields~\cite{Luis_JOpt2007} written here for zero time-delay ($\tau = 0$),
\begin{equation}
    |g_{n,m}^{(1)}(0)| = \left[ \frac{\text{Tr}{(\boldsymbol{\Gamma}_{n,m} \boldsymbol{\Gamma}_{n,m}^\dagger)}}{\text{Tr}{(\boldsymbol{\Gamma}_{n,n})} \text{Tr}{(\boldsymbol{\Gamma}_{m,m})} } \right]^{1/2},
\end{equation}
where the $2\times2$ correlation matrix is written,
\begin{equation}
    \boldsymbol{\Gamma}_{n,m} = \begin{pmatrix}
    \langle\psi_{+,n}^* \psi_{+,m} \rangle & \langle\psi_{+,n}^* \psi_{-,m} \rangle \\
    \langle\psi_{-,n}^* \psi_{+,m} \rangle & \langle\psi_{-,n}^* \psi_{-,m} \rangle
    \end{pmatrix}.
\end{equation}
Here, we have used the shorthand notation $\psi_\sigma(\mathbf{r} - \mathbf{r}_n) = \psi_{\sigma,n}$ and $\langle . \rangle$ stands for time average in simulation with random noise added at every time step. Specifically, a noise term is appended to the left hand side of Eq.~\eqref{eq.GPE} written $\dots + d\theta_\sigma(t)/dt$ where $\langle d\theta_\sigma(\mathbf{r},t) d\theta_{\sigma'}(\mathbf{r}',t') \rangle = 0$ and $\langle d\theta_\sigma(\mathbf{r},t) d\theta^*_{\sigma'}(\mathbf{r}',t') \rangle  = \eta \delta_{\sigma \sigma'} \delta(\mathbf{r}-\mathbf{r}') \delta(t-t')$ are the correlators of the complex-valued random Gaussian noise, and $\eta$ denotes its strength. Here, we are not concerned with the exact form of the noise which, in principle, should depend on the reservoir density.
\begin{figure}
    \centering
    \includegraphics[width = \linewidth]{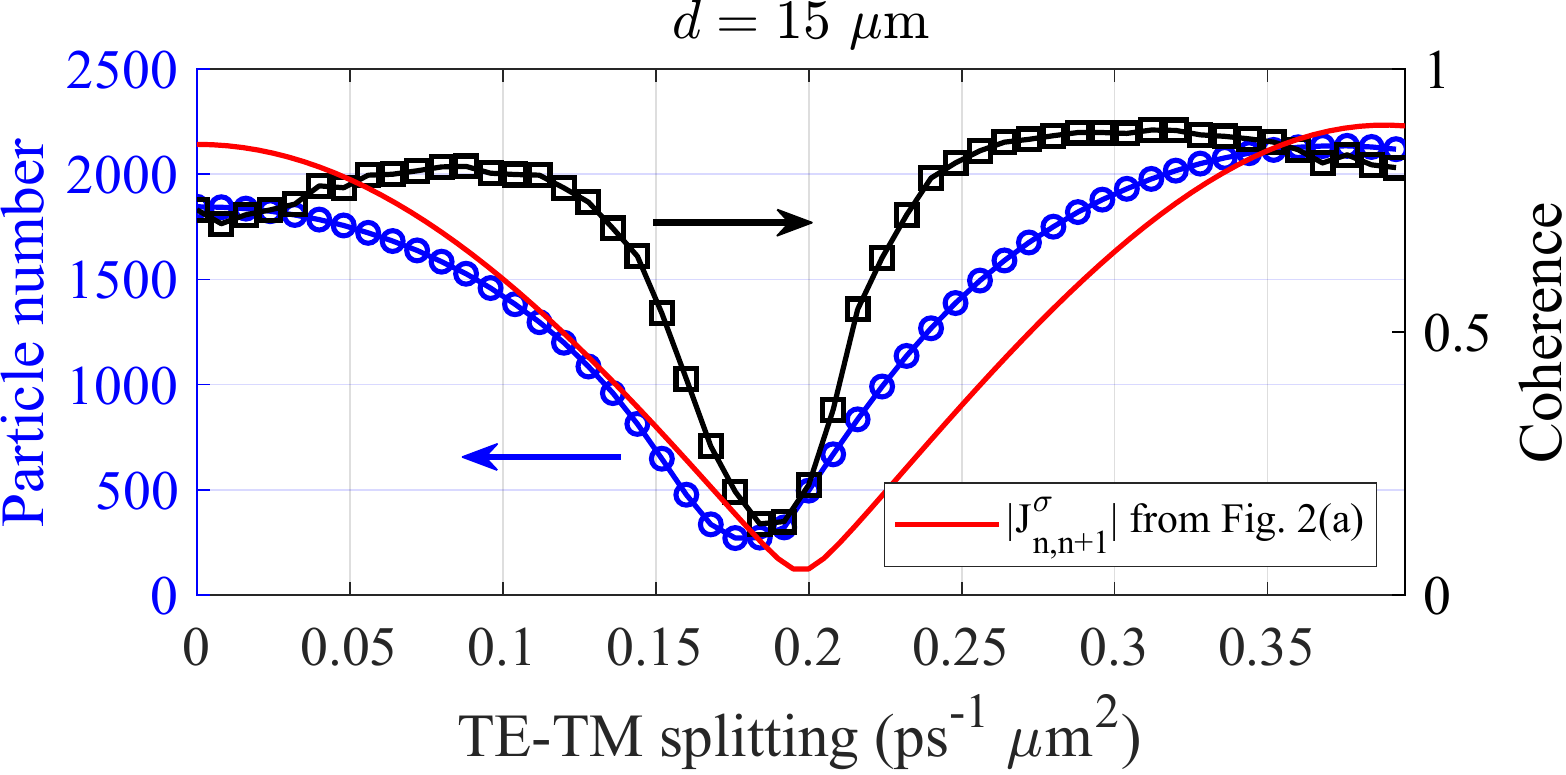}
    \caption{Total particle number (blue circles) and first order correlation function $|g_{-\mathbf{d}/2,\mathbf{d}/2}^{(1)}(0)|$ (black squares) for the simulated two condensate system as a function of TE-TM splitting. Both quantities display a minimum approximately matching the estimated minimum of their coupling strength (red solid line in arbitrary units) taken from the corresponding vertical line profile in Fig.~\ref{fig2}(a).}
    \label{fig4}
\end{figure}

Figure~\ref{fig4} shows the expected drop in particle number (blue circles) and coherence (black squares) when the TE-TM splitting is scanned, approximately coinciding with the minimum of nearest neighbor overlap $|J_{n,n+1}^\sigma|$ [red solid line (scaled)] extracted from a vertical line profile in Fig.~\ref{fig2}(a). The small mismatch can be understood from the fact that in the two-pump simulation the condensates have coupled together to form leaky standing waves with slightly shifted energies and momentum from the bare (single-pump) condensate. These results underline the reduced ability of the two condensates to synchronize and form a macroscopic coherent state at specific values of TE-TM splitting. 

\subsection{Three condensates}
\begin{figure}[t]
    \centering
    \includegraphics[width = \linewidth]{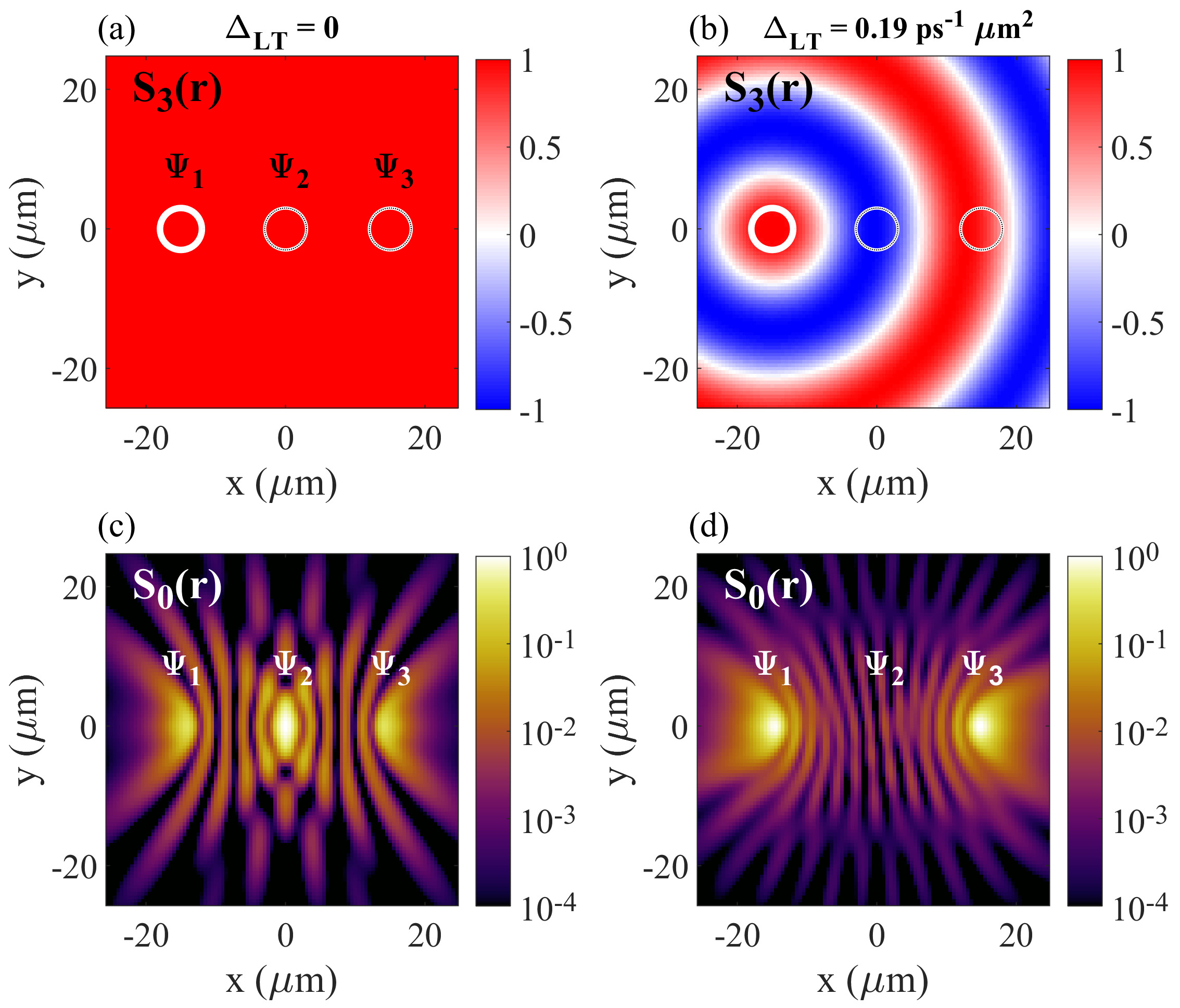}
    \caption{Real space $S_3$ Stokes component for only the left spot $\Psi_1$ pumped above threshold without TE-TM splitting (a) and with TE-TM splitting (b). Circles denote the location of the pumps used. Bottom panels show the lowest threshold steady state solutions in case of three equally $\sigma^+$ pumped condensate spots separated by $d = 15$ $\mu$m without TE-TM splitting (c) and with TE-TM splitting (d). Here the $S_0$ component is normalized in each case to more clearly show the resulting patterns.}
    \label{fig5}
\end{figure}

In Fig.~\ref{fig5} we show the possibility to use TE-TM splitting to allow dominant NNN interaction in a chain of three condensates. Just like in Fig.~\ref{fig3}, we first present the locations of the three pump spots (white circles) with only the left condensate pumped above threshold, without and with TE-TM splitting in Figs.~\ref{fig5}(a) and ~\ref{fig5}(b). Here, the middle pump spot is located in approximately the half-period of the $S_3$ precession whereas the right pump spot after a full-period.

Figure~\ref{fig5}(c) shows that in the absence of TE-TM splitting the interaction between NNs prevails and the lowest threshold solution corresponds to a central spot with dominant intensity due its reduced transverse losses and being strongly coupled to its two NNs. In contrast, when TE-TM splitting is tuned to make the NNN interactions strongest the lowest threshold solution is dramatically different as seen in Fig.~\ref{fig5}(d). The edge condensates strongly interact with each other and emit brightest, while the central condensate is suppressed. This can be interpreted as a type of \emph{spin-screening} where polaritons emitted from the edge spots convert their spins into the opposite component and thus pass through the central pump spot with minimal interference or scattering. It can be imagined as a waves from the edge condensates are "diving below'' the middle condensate, almost completely ignoring it's influence, and "resurfacing" at the opposite edge, carrying almost unperturbed information. It is worth noting that if the central pump spot were now excited with a $\sigma^-$ polarization the strong NN coupling would be restored.
\begin{figure}[t]
    \centering
    \includegraphics[width = \linewidth]{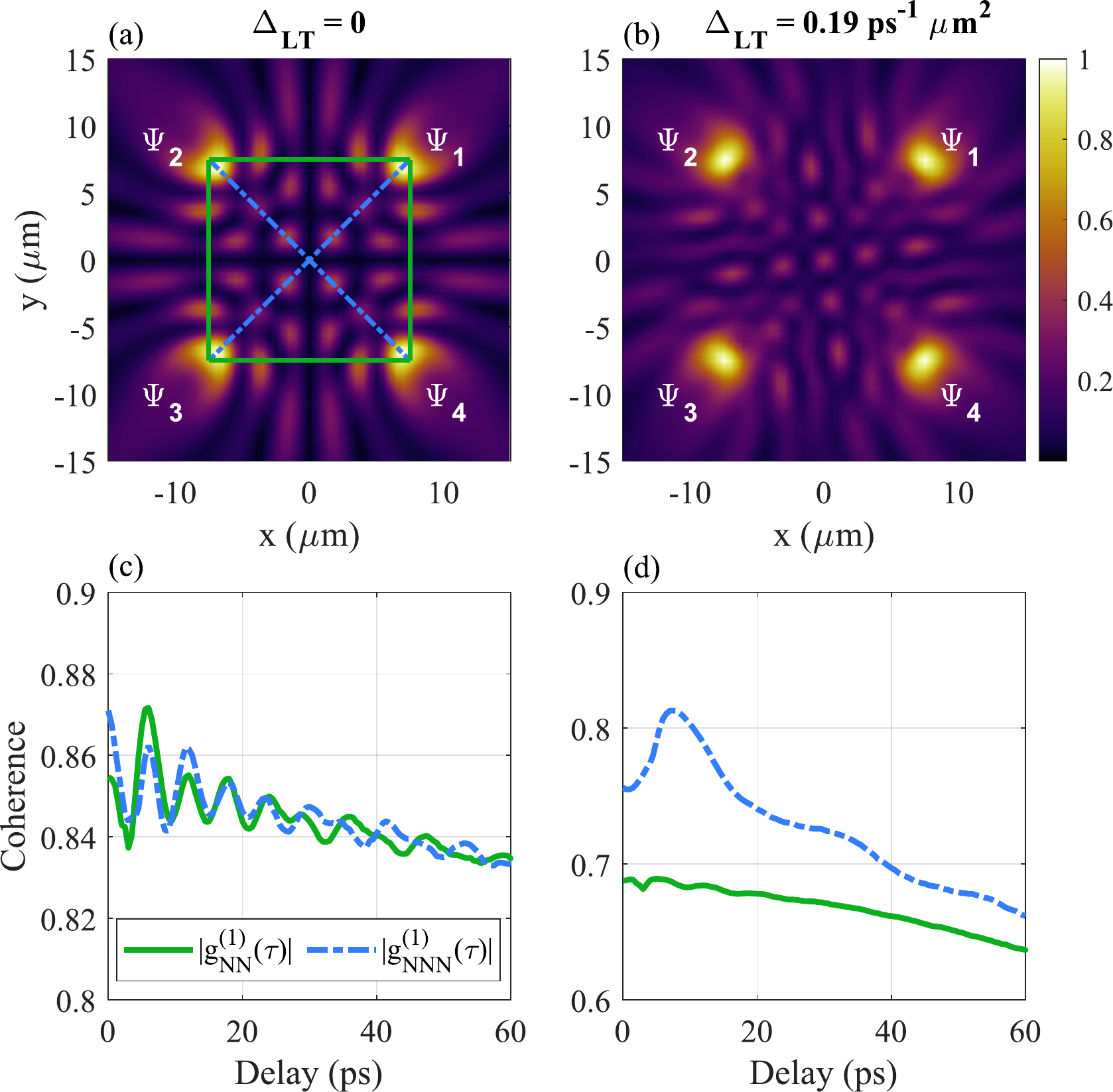}
    \caption{Time-averaged total density $S_0(\mathbf{r})$ in a square of pump spots with edge length $d = 15$ $\mu$m without (a) and with TE-TM splitting (b). Colored lines denote NNs (green) and NNNs (blue). Here, simulations are performed with stochastic noise present in the system. Calculated first order correlation function $|g^{(1)}_{1,m}(\tau)|$ between vertex pairs showing a dramatic change in coherence between NNs (green curve) and NNNs (blue curve) when going from zero (c) to finite TE-TM splitting (d). The $S_0$ component is normalized individually to more clearly show the resulting patterns.}
    \label{fig6}
\end{figure}

\subsection{Square of condensates}
The same effect can be used to modify the interaction in a square of four condensates~\cite{Tosi_NatComm2012, Alyatkin_PRL2020}. Figure~\ref{fig6}(a) shows the condensates time-averaged real space density without TE-TM splitting where each vertex has synchronized anti-phase with its neighbor with a clear dark node in the center of the square. In this section we have chosen to display the time-averaged $S_0$ due to small non-stationary dynamics in the condensate in this chosen geometric configuration [see oscillations in Fig.~\ref{fig6}(c)]. In the presence of noise, the first order correlation function as a function of time-delay $|g^{(1)}_{n,m}(\tau)|$ shows robust coherence between all corners of the square [see Fig.~\ref{fig6}(c)] indicating strong spatial coupling between all condensate pairs. 

The edges of the square form NNs with separation distance $d$ whereas the diagonals form NNNs with separation distance $\sqrt{2}d$ which allows us to choose a TE-TM splitting that maximizes the contrast between NNs and NNNs interaction strength similar to Fig.~\ref{fig5}. In this case the condensate density $S_0(\mathbf{r})$ displays a different parity structure with dominantly in-phase synchronization and a bright antinodal spot in the square center [see Fig.~\ref{fig6}(b)]. As expected, with NNs interactions weakened the first order correlation function displays stronger coherence now between the diagonals [blue dot-dashed curve in Fig.~\ref{fig6}(d)].

\section{Conclusions}
We have proposed and analyzed a method of making next-nearest-neighbor interactions stronger than nearest-neighbor interactions in networks of ballistic exciton-polariton condensates. Photonic spin-orbit coupling, arising from inherent splitting of the cavity TE and TM modes, results in rapid precession of the polariton pseudospin propagating in the cavity plane, modifying the interference between neighboring condensates. The spatial overlap between neighboring condensates can therefore be tuned by simply combining the geometry of the condensate network with the strength of the TE-TM splitting such that nearest-neighbors become effectively \emph{spin-screened} while next-nearest-neighbors still interact strongly. This phenomena leads to several interesting effects such as completely new set of low-threshold condensation modes [Fig.~\ref{fig5}], lower spatial emission [Fig.~\ref{fig3}], and reduced coherence between nearest-neighbors [Fig.~\ref{fig4} and Fig.~\ref{fig6}]. Furthermore, our method and findings are not exclusive to the spin-orbit-coupling from photonic TE-TM splitting but can also be used with the more familiar Rashba or Dresselhaus spin-orbit-coupling~\cite{Galitski_Nature2013}.

Notably, our method opens the possibility to study non-planar graph problems in networks of polariton condensates. The famous non-planar $K_5$ graph, which is used to determine graph non-planarity through Kuratowski's theorem, can be easily constructed in the polariton system with five lasers forming a pentagram. Polariton networks then offer new perspectives on simulating XY phases~\cite{Berloff_NatMat2017} in complex graphs in the optical regime. As an example, non-planarity in the well-known Ising model renders it NP-complete~\cite{Barahona_IOP1982}, making it impossible to solve efficiently, which underpins the importance of designing physical alternatives to simulate such complex systems. Such a prominent alternative are Rydberg atoms in optical tweezers with tunable neighbor interactions~\cite{Browaeys_NatRevPhys2020, Ebadi_Nature2021} that permit exploration of exotic quantum phase diagrams. Polariton condensates, with our method, can thus offer insight on hard-to-reach phases in the semiclassical regime describing interacting driven-dissipative nonlinear oscillators.

Our study takes a step forward in development of optically programmable networks of interacting nonequilibrium quantum gases. It offers new perspectives on utilizing photonic spin orbit coupling to directly manipulate the spatial degrees of freedom in polariton condensate networks and lattices which have experienced serious advancements in the past years~\cite{Amo_ComRenPhys2016, Berloff_NatMat2017, Ohadi_PRB2018, Zhang_NanoScale2018, Hakala_NatPhys2018, Alyatkin_PRL2020, Pickup_NatComm2020, Ballarini_NanoLett2020, Jayaprakash_ACSPho2020, Topfer_Optica2021, Scafirimuto_CommPhys2021, Alyatkin_NatComm2021, Pieczarka_Optica2021}. Our method can permit access towards unexplored regimes of synchronicity in large-scale spinor polariton fluids.  

The data that support the findings of this study are openly available from the University of Southampton repository~\cite{data}

\section{Acknowledgements}
The reported study was funded by the Russian Foundation for Basic Research (RFBR), project number 20-02-00919. The authors acknowledge the support of the UK’s Engineering and Physical Sciences Research Council (grant EP/M025330/1 on Hybrid Polaritonics), and European Union’s Horizon 2020 program, through a FET Open research and innovation action under the grant agreement No. 899141 (PoLLoC). H.S. acknowledges the Icelandic Research Fund (Rannis), grant No. 217631-051.

\appendix

\section{Reservoir spin-relaxation} \label{app1}
In this section we provide a short analysis and discussion on the effects of a simple spin-relaxation mechanism in the exciton reservoir. For simplicity, we will work in the linear regime (i.e., below threshold $|\psi_\pm|^2 \simeq 0$) and therefore spatial degrees can be omitted. Equation~\eqref{eq.Res} is now written,
\begin{equation}
    \frac{d X_{\pm} }{dt} = -\Gamma X_\pm + \Gamma_s(X_\mp - X_\pm) +  P_\pm,
\end{equation}
where $\Gamma_s$ describes the rate of spin conversion between the exciton components. If we parametrize the pump as $\mathbf{P} = (P_+,P_-)^\text{T} = P_0 (\cos^2{(\theta)}, \sin^2{(\theta)})^\text{T}$ the steady state solution can be written,
\begin{equation}
    \mathbf{X} = \frac{P_0}{\Gamma(\Gamma + 2\Gamma_s)} \begin{pmatrix} \Gamma \cos^2{(\theta)} + \Gamma_s \\
    \Gamma \sin^2{(\theta)} + \Gamma_s \end{pmatrix}.
\end{equation}
If the pump is $100\%$ $\sigma^+$ polarized ($\theta=0$) one obtains
\begin{equation}
    \frac{X_+}{X_-} = \frac{\Gamma + \Gamma_s}{\Gamma_s}.
\end{equation}
Since the reservoir is proportional to the pump power, we see that the inequality $P_\text{th} < P_0 < P_\text{th} (\Gamma + \Gamma_s)/ \Gamma_s$ should be satisfied (where $P_\text{th}$ is the threshold power for the $\psi_+$ polaritons) in order for the $\psi_-$ polaritons to remain uncondensed at the pump spots.

The upper bound on the pumping power is absolute for strongly spin polarized condensates at their pump spots. However, the complex interplay of reservoir spin-relaxation, shape of the pumping potential, and TE-TM splitting might also lead to elliptically polarized condensate solutions with lower threshold than $P_\text{th} (\Gamma + \Gamma_s)/ \Gamma_s$. Nonetheless, the above short analysis serves as a good indicator on the range of valid pump powers to produce our study's findings in experiment. 
\begin{figure}[b]
    \centering
    \includegraphics[width = \linewidth]{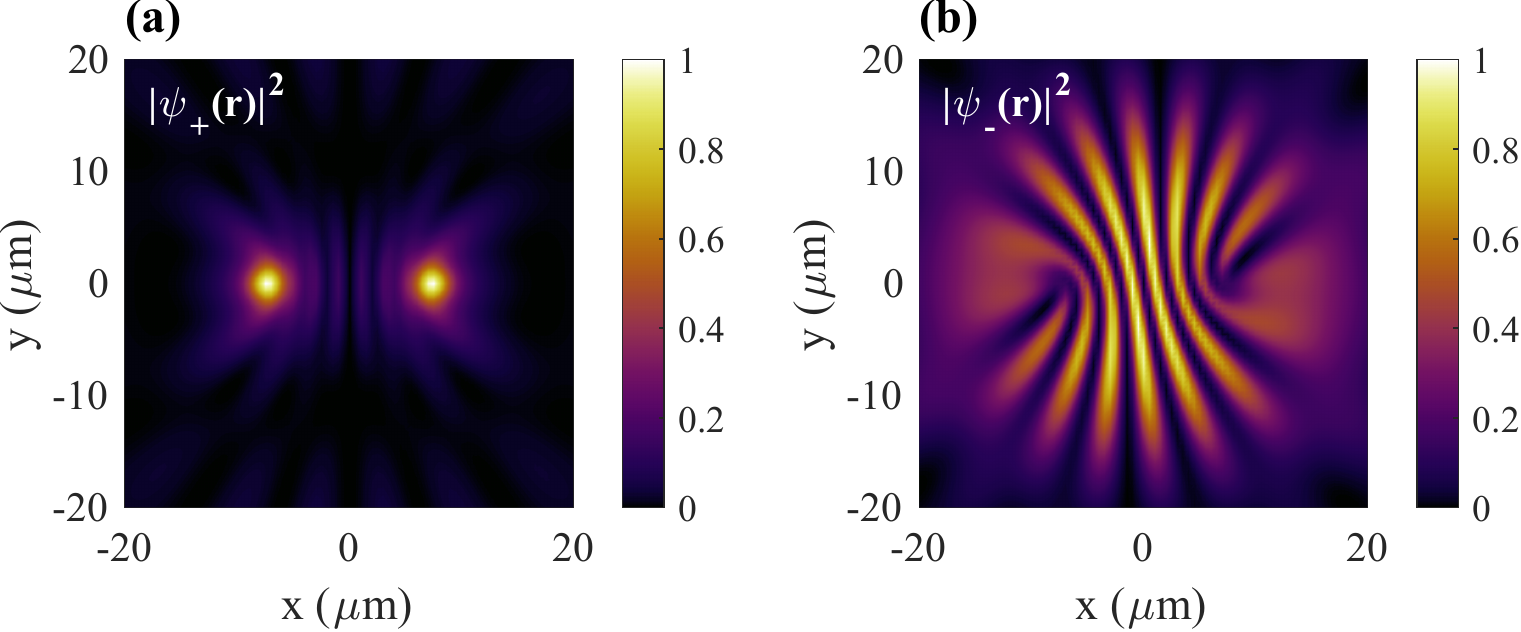}
    \caption{Spin up (a) and spin down (b) condensate densities from Fig.~\ref{fig3}(d) showing the mirror symmetric standing wave pattern in the pumped $\psi_+$ polaritons and a chiral pattern in the converted $\psi_-$ polaritons due to SOC. Each panel is normalized independently.}
    \label{fig7}
\end{figure}

\section{Chiral density distribution} \label{app2}
In this section we break down the $S_0(\mathbf{r})$ Stokes parameter in Fig.~\ref{fig3}(d) into its two components $|\psi_\pm(\mathbf{r})|^2$ and plot them in Fig.~\ref{fig7}. The results show a symmetric standing wave pattern for the pumped spin-up polaritons but a chiral pattern for the spin-down polaritons. This chirality naturally reverses if we instead pump the spin-down polaritons with $\sigma^-$ light, and cancels out if we pump both spin components equally with linearly polarized light.

\end{document}